\documentclass[preprint,aps]{revtex4}
\usepackage{graphicx}
\usepackage{dcolumn}
\usepackage{rotating}
\newcommand{\bg}{\begin{equation}}
\newcommand{\ed}{\end{equation}}

\begin{document}

\title {Transport in Molecular Junctions with Different Metallic Contacts}

\author{John W. Lawson}
\email{John.W.Lawson@nasa.gov}
\author{Charles W. Bauschlicher, Jr.} 
\email{Charles.W.Bauschlicher@nasa.gov}

\affiliation{Mail Stop 269-2\\
Center for Nanotechnology \\
NASA Ames Research Center\\
Moffett Field, CA 94035 }

\begin{abstract}
{\normalsize
Ab initio calculations of phenyl dithiol connected to Au, Ag, Pd, and Pt electrodes
are performed using non-equilibrium Green's functions and density functional 
theory.  For each metal, the properties of the molecular junction are considered
both in equilibrium and under bias.  In particular, we consider in detail charge
transfer, changes in the electrostatic potential, and their subsequent effects
on the IV curves through the junctions.  Gold is typically used in molecular
junctions because it forms strong chemical bonds with sulfur.  We find however
that Pt and Pd make better electrical contacts than Au.  The zero-bias conductance is
found to be greatest for Pt, followed by Pd, Au, and then Ag.
}
\end{abstract}

\maketitle
\section{Introduction}

Interest in electrical conduction in molecules has been spurred in recent
years by many experimental results where transport in individual molecules
was measured.  A number of interesting and potentially technologically useful
phenomena have been catalogued from switching \cite{exp1}, nondifferential
resistance \cite{exp2}, and transistor action \cite{exp3} to more exotic
behavior such as Kondo physics \cite{exp4} and vibronic effects \cite{exp5}.
In parallel with these developments has been theoretical work, ranging from
semi-empirical theories \cite{dattaeht} to ab initio formalisms based on
non-equilibrium Green's functions (NEGF) and density functional theory (DFT)
\cite{xr1,t2,t3,t4,t5,t6,t7,t8,t9,t10}.  Many open questions remain including
the large discrepancy between theoretical and experimental current-voltage
IV curves as well as attempts to obtain more accurate experimental
characterization of the junctions. 

Phenyl dithiol (PDT) attached to Au electrodes has become an important
prototypical system.  This is due in part to important early
experiments\cite{reed}, but is also due to its accessibility to a number of
high-powered theoretical tools.  Experimental interest in Au contacts is
largely one of convenience, since Au is known to make good chemical contact
with the thiol end groups.  Whether this results in an optimal electrical
contact is less clear.  It is important therefore to examine the conduction
properties of PDT connected to other metals in an effort to find
the combination with optimal performance characteristics.

In this paper, we compare in detail the transport properties of PDT with Ag, Au, Pd,
and Pt contacts.  Experimental investigations of molecular junctions with
electrodes other than gold have included molecular hydrogen attached to Pd
and Pt \cite{h2} as well as more complicated organic molecules in contact
with different metals \cite{org}.  To our knowledge, there has not been an
experimental study of PDT with non-gold contacts.  Early theoretical analysis
of these systems has appeared \cite{sem}.  However, a simpler formulation was
utilized than what we use, especially with respect to the description of the
contacts, which is our main object of study.  Our results show significant
differences from that previous work.  Yaliraki et al \cite{yakrat} using
a non-self-consistent method considered Ag in addition to Au.  They found Ag to
be a worse conductor than Au consistent with our results.  Di Ventra et al.
\cite{divent} considered Al contacts in addition to Au, finding the Al junction
to have better conduction characteristics.  In addition to current-voltage (IV)
characteristics, we consider the role of charge transfer and changes in
electrostatic potential due to formation of the contacts, both at equilibrium
and under bias.

In general, the tunneling current through a molecular junction depends
on the electronic structure of the junction in the vicinity of the Fermi
level.  We found that junctions with different metal electrodes result in
qualitatively different conduction characteristics.  We employed a first
principles method based on NEGF and DFT.
This allowed us to consider the interplay of
issues related to charge transfer and band lineup due to the formation of
contacts under equilibrium condition as well as nonequilibrium transport
through the junction under bias.

\par
\section{Method}

We utilized the methodology developed by Xue, Datta, and Ratner \cite{xr2}.
The NEGF code has been interfaced with Gaussian03, which allows us
access to the full suite of basis sets and functionals offered by Gaussian03 to
describe our systems.  Details of our calculational approach and methodology have
been reported elsewhere \cite{xr2,xue031,xue032,xue04,xueth,cwb1}.  We recall the
most salient features briefly for completeness.

Clusters of atoms from each metal contact are added to the ends of the molecule
to form an ``extended molecule".  The extended molecule is then further connected
to additional atoms in the electrodes.  This coupling is implemented through
a Green's function
approach
\bg
G(E)=1/(ES-F-\Sigma_L(E)-\Sigma_R(E))
\ed
where $F$ is the Fock matrix of the extended molecule, $S$ is the overlap matrix,
and $\Sigma_L(E),\Sigma_R(E)$ are self-energies that define the couplings to the
contacts.  The self-energies are defined as
\bg
\Sigma(E)=\tau^{T}g_s(E)\tau
\ed
where $\tau$ is a matrix that gives the couplings between the extended molecule and
atoms in the contacts and $g_s(E)$ is a matrix representing the surface Green's function
for a semi-infinite bulk metal.  The density of states (DOS) for the metal
surfaces can be recovered by $DOS(E)=-Im(g_s(E))/2\pi$.

From the Green's function, the density matrix is calculated as
\bg
\rho = \frac{1}{2 \pi i} \int_C dE G(E)
\ed
and, thus within a DFT framework, a self-consistent field (SCF) procedure can be devised.
Self-consistent methods are necessary to describe charge transfer effects
correctly.  Since $G(E)$ has a non-trivial energy dependence through $\Sigma(E)$,
the integral is performed using a numerical complex contour technique.

Once a self-consistent Green's function has been obtained, the transmission
function can be evaluated using the Landauer relation
\bg
T(E)=Tr[\Gamma_L(E) G(E) \Gamma_R(E) G^{\dagger}(E)].
\ed
where $\Gamma(E)=i(\Sigma(E)-\Sigma(E)^{\dagger})$ is the coupling function that 
gives information about the quality of the contact between the molecule and the
electrodes as well as information about the density of states in the bulk
available for current transmission across the junction.
The transmitted current can be calculated at zero temperature as
the integral of the transmission function
\bg
I=\frac{2e}{h}\int^{E_f+V/2}_{E_f-V/2}T(E)dE
\ed
in a energy window of width $V$ around the Fermi energy $E_f$.  Generalizing
this result to finite temperature is straightforward.

The self-energy $\Sigma(E)$ contains not only geometric and electronic information
about the contacts, but also defines the strength of the coupling between the
molecule and the electrode. The self-energy also acts as an energy dependent,
complex-valued potential.  Its effect in modeling the contacts is to shift
and broaden the energy levels of the molecule, determine the density of states
available to be transmitted from the metal, and give the strength of the coupling
to the molecule.  All of these factors affect the transmission spectra of the junction.
A comprehensive discussion of the NEGF approach to transmission is given by
Datta \cite{datta}.

Evaluation of $\Sigma(E)$ is made tractable by the fact that elements of $\tau$
are nonzero for only a distance of several atomic layers from the extended molecule.
In this work, the $\tau$ matrices are constructed from tight binding (TB) parameters
\cite{papa} where nine orbitals per Au atom have been used.  The TB parameters 
were obtained from fits to first principles band structure calculations.
The surface Green's function $g_s(E)$ for the semi-infinite metals is constructed 
using a recursive procedure \cite{flores}.  The Green's function is decomposed
into a sum
\bg
g_s(\vec{R},E)=\sum_{\vec{k}} g_{\vec{k}}^0(y,E) e^{i\vec{k} \cdot \vec{r}}
\ed
where $g_{\vec{k}}^0$ are Fourier components in the first principal
layer parallel to the surface and $\vec{r}$ and $\vec{k}$ are vectors 
parallel to the surface.  Each $g_{\vec{k}}^0$ is coupled
successively to components $g_{\vec{k}}^n$ residing in the $n^{th}$ principal layer
deeper in the bulk.  This process is continued
until the surface components $g_{\vec{k}}^0$ converge.  For computational expediency,
tight binding parameters are used to build the coupling matrices to start the
iteration.  This is the principle approximation for constructing $\Sigma(E)$. 
There are no free parameters in this formalism.

\section{Model}
A potential difficulty of our approach is the fact that the extended 
molecule is described with DFT while the self-energies $\Sigma(E)$, which represent
the wider electrodes, are constructed using tight binding parameters.
This mismatch of microscopic descriptions may affect the quality of
the contact, and lead, for example, to spurious reflections at the
boundary of the two regions.  To test this possibility, we examined
transmission though linear atomic chains where the chains and electrodes
were composed of the same metallic elements.  In these cases, we expect
an especially good contact to result and also expect the transmission to
approximate the theoretical maximum given by the quantum of conductance.

\begin{figure}
\begin{center}
\resizebox{150mm}{!}{\includegraphics{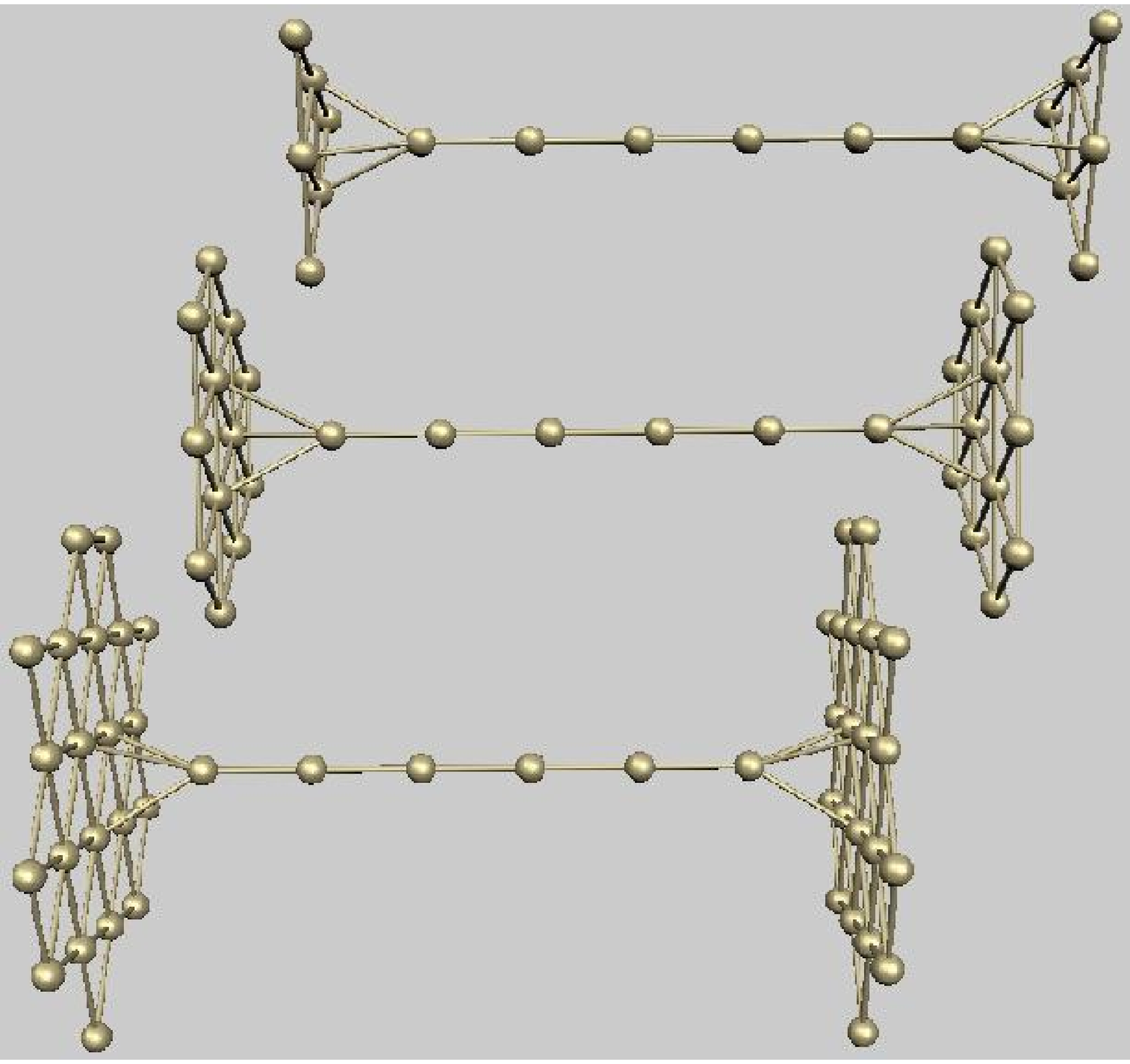}}
\caption{
\label{f1}
(Color online) Extended quantum point contact (QPC) composed of a linear chain of
six metal atoms.  The chain plus clusters of size 6, 12, and 21 
surface atoms defines the ``extended molecule" for these systems.
}
\end{center}
\end{figure}

\begin{figure}
\begin{center}
\resizebox{150mm}{!}{\includegraphics{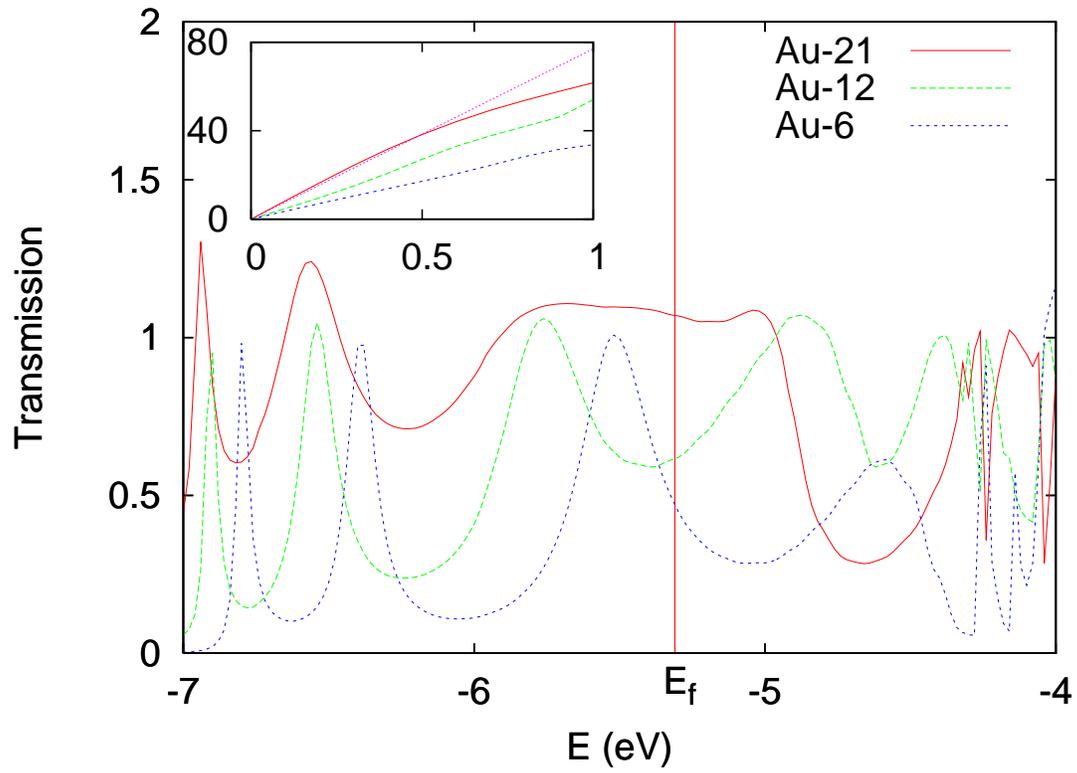}}
\caption{
\label{f2}
(Color online) Transmission spectrum for linear Au chain with Au electrodes.
Inset is corresponding IV curve where the upper dotted curve shows
the theoretical maximum given by the quantum of conductance.
Transmission increases with cluster size.
}
\end{center}
\end{figure}

In particular, we considered linear chains of six metallic atoms for Au, Pt,
Pd, and Ag.  The geometries considered are shown in Fig.~\ref{f1}.
Atoms in the first surface layer are added explicitly to the
extended molecule and are then included in the self-consistent field calculations.
Additional atoms in the both the first and second layers also coupled to
the extended molecule through the self-energies, although their orbitals
remain fixed during the SCF.
The Fermi energies used were $E_f=-5.31 eV$ for Au, $E_f=-5.6 eV$
for Pd, $E_f=-5.93 eV$ for Pt, and $E_f=-4.74 eV$ for Ag.  The interatomic
spacings within the chains were optimized and bulk positions were used for
the clusters attached at each end.  Equilibrium transmission spectra and IV
characteristics were calculated for the different metals with clusters of
6, 12, and 21 atoms included on each side on the extended molecule.
In Fig.~\ref{f2}, we show results for
gold which are representative.  The main graph is the transmission spectra
while the inset is the IV curves from zero to one volt.  In the IV curve,
the upper dotted line gives the theoretical maximum.  Clearly, as the size of the
cluster increases, there is a significant increase in the transmission especially
near the Fermi level.   The results for the 21 atom cluster 
are very close to the quantum of conductance.  There is a small 
overestimation of the transmission near the Fermi level.  This is an
artifact of the TB approximation of $\Sigma(E)$.  Based on these results,
we used the larger clusters in subsequent calculations. 

\begin{figure}
\begin{center}
\resizebox{150mm}{!}{\includegraphics{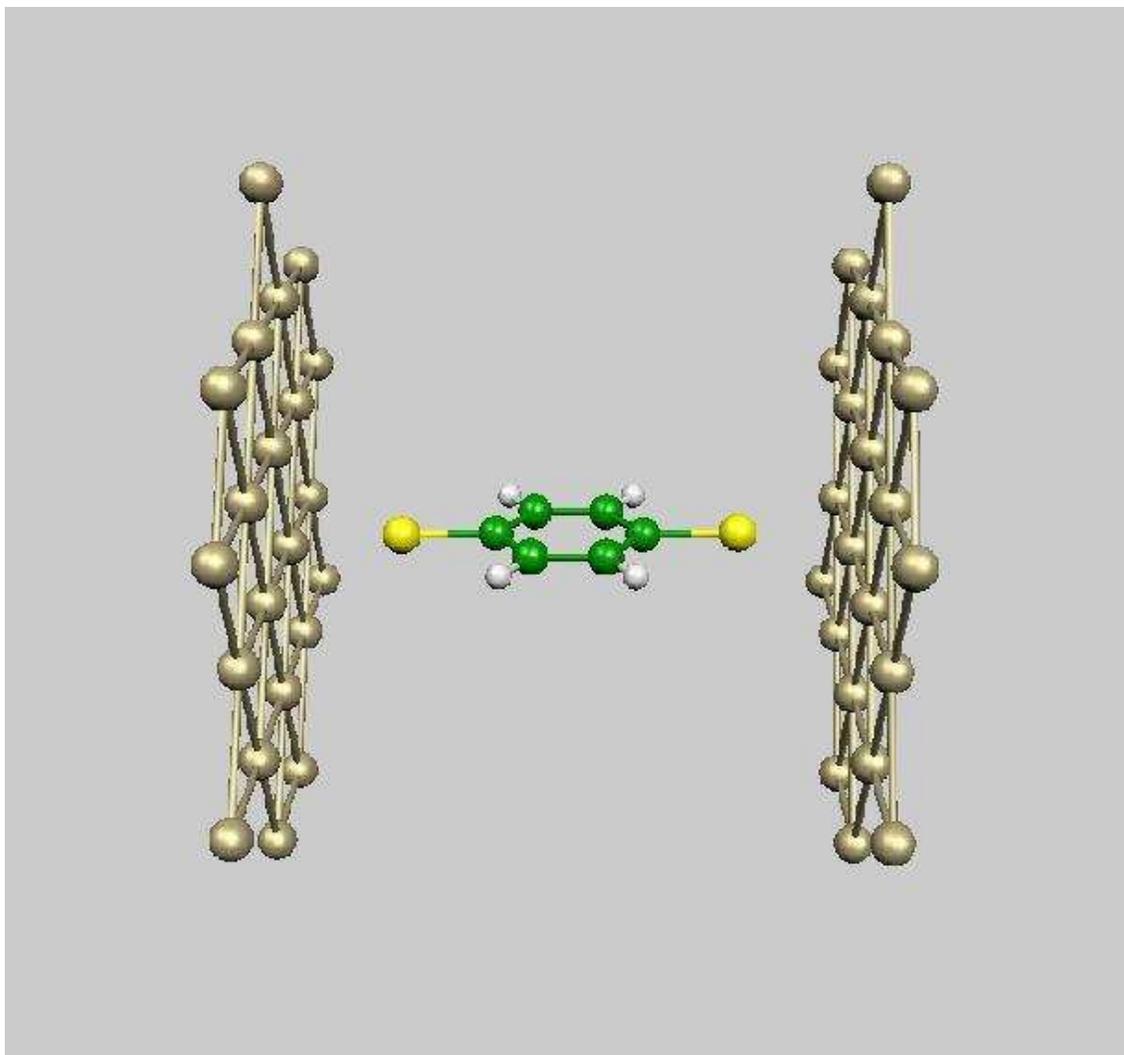}}
\caption{
\label{f3}
(Color online) Extended molecule with phenyl dithiol (PDT) with clusters 21 metal
surface atoms on each end. 
The extended molecule is coupled to an additional 63 atoms in the first
and second layers and to the semi-infinite bulk through the self-energies.
}
\end{center}
\end{figure}

The junctions we considered consist of a single molecular fragment,
PDT, $-S-C_6H_4-S-$ bonded to two parallel, semi-finite metallic electrodes.  
It should be noted that this is a non-periodic calculation.  Therefore, we model a 
single molecule, rather than an array.
The electrodes are represented by clusters of 21 atoms taken from the first surface
layer and included into the extended molecule and therefore into the self-consistent
field calculation.  An additional 63 atoms from both the first and second layers
are coupled to the extended molecule through $\Sigma(E)$.
However, the orbitals for these atoms do not
relax self-consistently, but are fixed to their TB values.
The geometry of the extended
molecule used is shown in Fig.~\ref{f3}.  The axis of the molecule is parallel
to the cartesian $y$-axis while the Au surfaces run parallel to the $xz$-plane.
The geometry of PDT was optimized with H atoms attached to the S, and for each
metal we optimized the metal-S bond length after removing the terminal H atoms.
It is important to note that to determine the bond lengths, we optimized the energy
of the entire junction including the effects of the self-energies.  We found
significant variation using simpler models, differing basis sets, or by neglecting the
self-energies.  The results were $d=2.379 \AA$ for gold, $d=2.569 \AA$ for platinum,
$d=2.776 \AA$ for palladium, and $d=2.876 \AA$ Silver.  The sulfur were positioned
over the hollow of the $<111>$ surface and the molecule was oriented at 90
degrees to the surface.  This orientation has been shown in previous work
\cite{cwbar}, and by others, to be the low energy configuration.

The electronic cores of the molecule (C,S) were replace by compact effective
potentials (CEP).  The valence electrons were described by the CEP-121G basis
set where we have added polarization functions to the carbon and hydrogen and 
diffuse and polarization functions to the sulfur.  Previous work has shown
\cite{cwb1} that inclusion of polarization/diffuse functions has an important
impact on the results.  In particular, larger currents were observed for the
basis set CEP121+G* as the sulfur was able to make a better contact
with the electrodes \cite{cwb1}.

For the metallic atoms, we utilize the LANL1 pseudopotential and a
``reduced/optimized" minimal basis set.  For Au and Pt, this means removing
the most diffuse s,p, and d Gaussian primitives from basis functions, and
for Pd and Ag, removing only the most diffuse s and p.  Previous work
\cite{xue031,cwbxue} as shown that elimination of the most diffuse functions
on the metal atoms reduces unphysical super-charging of the contacts and also
improves convergence.  Any residual charging
can be further reduced to zero by optimizing the the d-electron basis set
(in particular, the most diffuse $d$ functions) and also by
adding a small field (~5-10E-4 a.u.) to the contacts.  This has the effect
analogous to periodic calculations where the Hartree potential is forced to
maintain the bulk values deep inside the electrodes.

All electronic structure calculations were performed at the DFT level using Gaussian03.
Green's function calculations were performed using the code described previously.
We mainly used the hybrid
functional B3PW91 \cite{becke,pw91} due to its greater accuracy, although we
also used other functionals, such as BPW91, for purposes of comparison.
Previous work has shown that hybrid functionals tend to reduce the magnitude of
the current reflecting the effects of exchange \cite{cwb1}.
All calculations are done at zero temperature.

\par
\section{Results and Discussion}

\begin{figure}
\begin{center}
\resizebox{150mm}{!}{\includegraphics{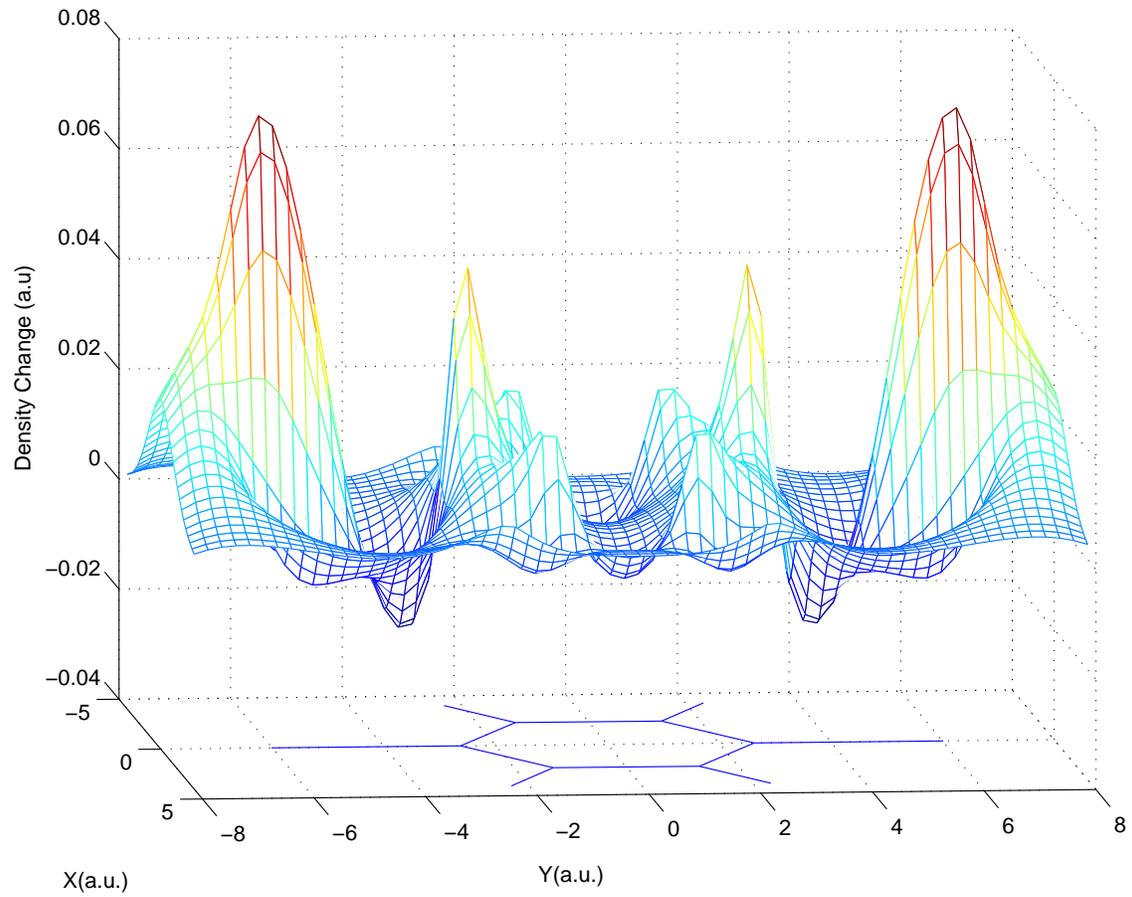}}
\caption{
\label{f4}
(Color online) Spatial distribution of the charge density change upon formation of the
contacts for Ag junction.  
}
\end{center}
\end{figure}

We began by considering the zero bias equilibrium situation to examine the
effects of the binding of the molecule to the metallic electrodes.  To compare
the degree of charge transfer for the different electrodes, we considered the
spatial distribution of the charge redistribution resulting from the junction formation.
As an example, we show in Fig.~\ref{f4} the spatial profile of charge redistribution
for the Ag junction that has the largest transfer.  The figure is an XY plot,
where the results have been averaged in the z-direction, of the difference between
the charge density of the device at equilibrium and the contact plus the bare molecule
densities taken alone.  We see that the binding of the sulfur to the metallic Ag
atoms results in charge transfer to the sulfur and the adjacent carbon.  We further
note a depletion of charge in the region between the S-C atoms which indicates a
weakening of those bonds.

\begin{figure}
\begin{center}
\resizebox{150mm}{!}{\includegraphics{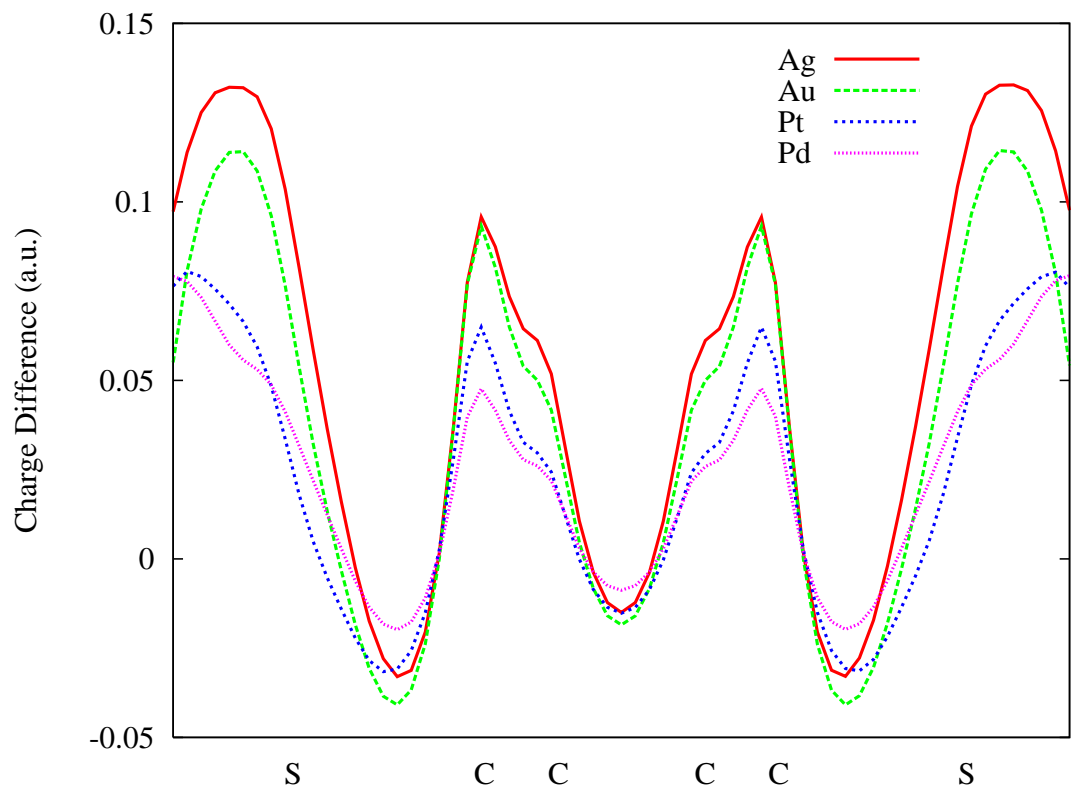}}
\caption{
\label{f5}
(Color online) Comparison of the equilibrium charge density change along the axis
of the molecule.  The largest transfer is for Ag with Pd have the smallest.
}
\end{center}
\end{figure}

For all four metals, we see a similar pattern, but with different magnitudes.
In Fig.~\ref{f5}, we further averaged over the x-direction to compare charge
transfer along the axis of the molecule between the different
junctions.  We indicate the $y$-coordinate of the molecular atoms on the horizontal
axis.  We see that Ag has the largest charge transfer following by Au, Pt, and then Pd.

\begin{figure}
\begin{center}
\resizebox{150mm}{!}{\includegraphics{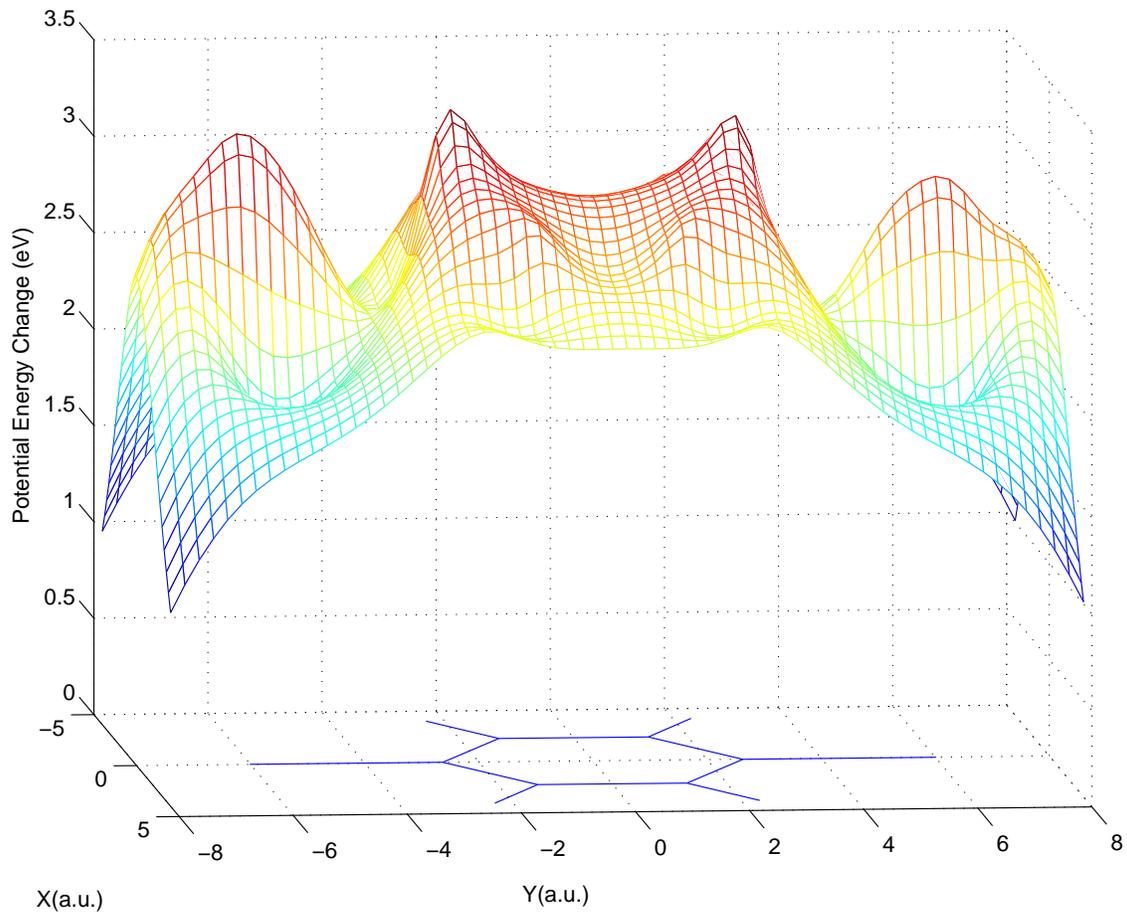}}
\caption{
\label{f6}
(Color online) Spatial profile of the electrostatic potential energy change across the Ag
junction.  Potential barriers form near the sulfurs and adjacent carbons.
}
\end{center}
\end{figure}

\begin{figure}
\begin{center}
\resizebox{150mm}{!}{\includegraphics{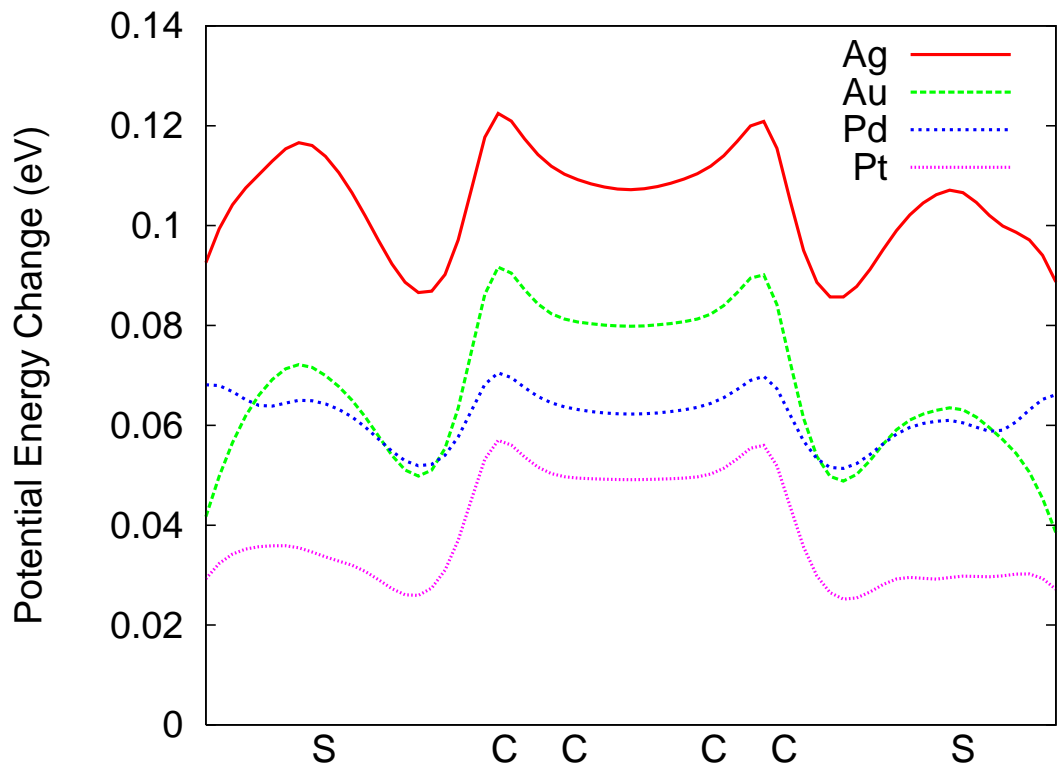}}
\caption{
\label{f7}
(Color online) Comparison of the equilibrium electrostatic potential energy change
along the axis of the molecule due to contact formation.
The barrier for Ag is twice that of Au and four times that of Pt. 
}
\end{center}
\end{figure}

To gain further insight into junction formation, we also examine the
corresponding change in the electrostatic potential energy.  In Fig.~\ref{f6},
we show the full spatial profile for Ag, the metal with the largest
redistribution.  For the potential profile, we show a cross-section of the
potential taken in the XY plane ($z=0$).  The z-direction has not been averaged
in order to display the potential energy the electron actually sees.
Potential barriers are formed on the sulfur atoms as well as on the interior carbons.
These barriers are relevant for tunneling electrons
through the junction under bias.  In Fig.~\ref{f7}, we compare the potential
energy redistributions along the axis of the molecule.  We see that Ag has the
largest barriers to overcome for electron transport and Pt the smallest.  On
this basis, we might guess that Pt would make the best conductor and Ag the worst.
We also note that Ag has a barrier twice that of Au and four times that of Pt.

\begin{figure}
\begin{center}
\resizebox{150mm}{!}{\includegraphics{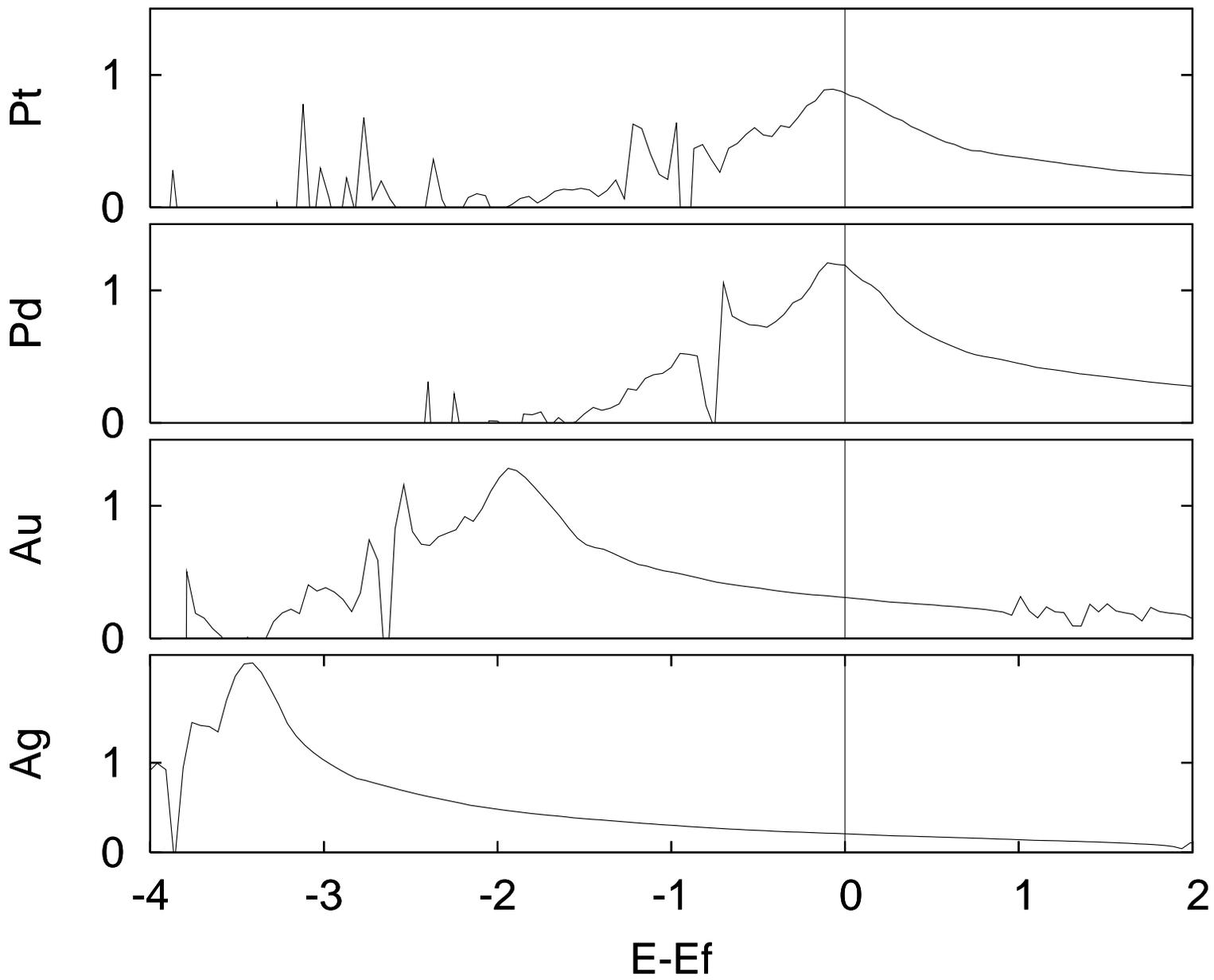}}
\caption{
\label{f8}
DOS for the electrode metals calculated from the surface Green's function.
}
\end{center}
\end{figure}

The self-consistent change in potential affects the electronic structure of
the molecule, leading to a lineup of molecular states with the continuum of
states residing in the electrodes.
In order for there to be transmission, there must be finite density of states
in the contacts to be transmitted.  The surface DOS can be calculated from the
surface Green's function $g_s(E)$ as $DOS(E)=-Im(g_s(E))/2\pi$.
We compare the DOS for the the metals in Fig.~\ref{f8}.
The energy scale is given relative to the individual Fermi energies.  We see
from the figure that in all cases there are states available at the Fermi level
for transport.  We notice in particular that while for Au and Ag, the DOS is
rather flat, for Pt and Pd, the DOSs are rapidly increasing near their Fermi energy.

\begin{figure}
\begin{center}
\resizebox{150mm}{!}{\includegraphics{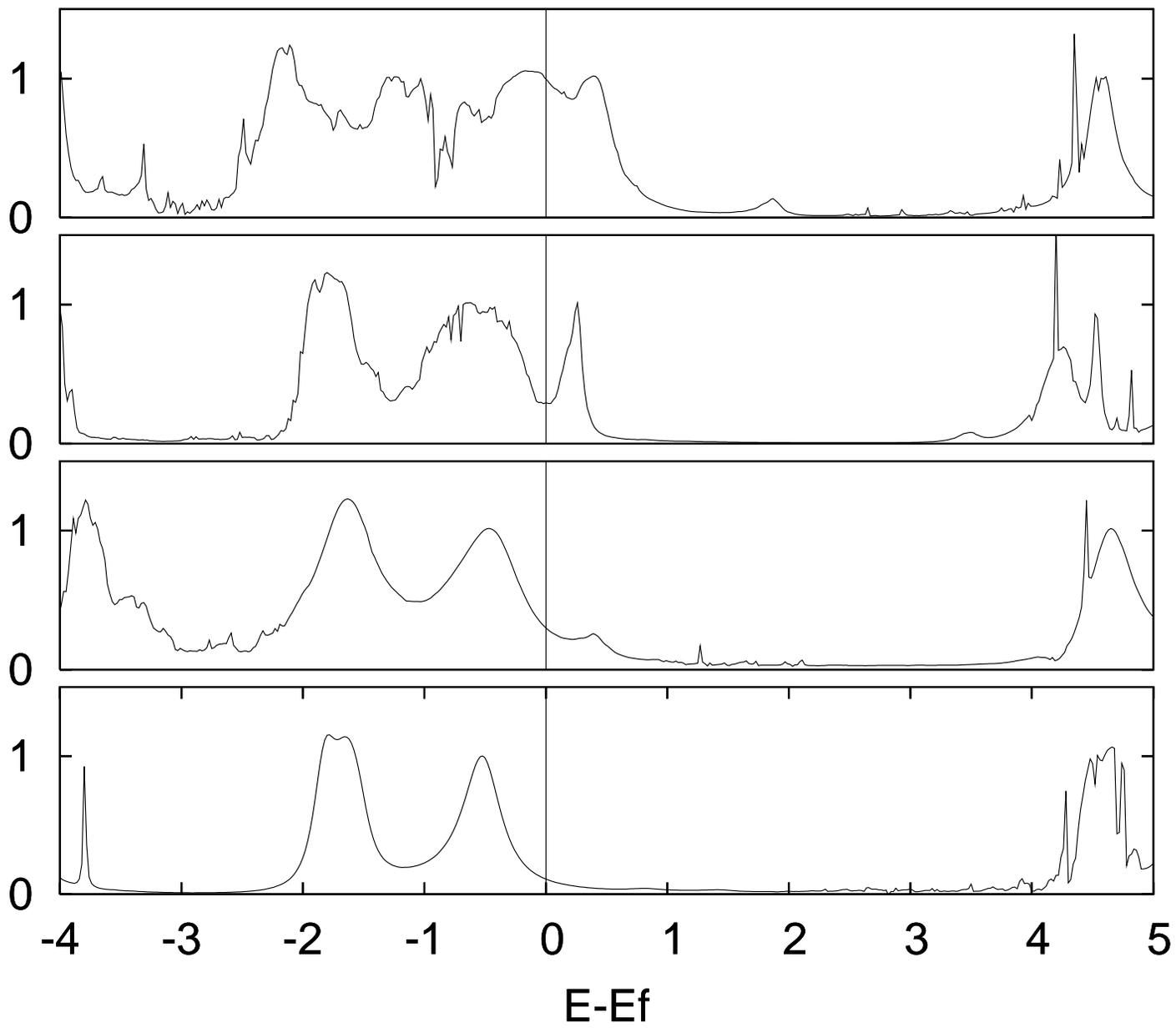}}
\caption{
\label{f9}
Zero-bias transmission spectra for the four junctions. 
}
\end{center}
\end{figure}

The self-consistent change in potential affects the electronic structure of
the molecule, leading to a lineup of molecular states with the continuum of
states residing in the electrodes.  The affect of this lineup as well as the level
broadening due to coupling the contacts can be seen directly in the transmission
spectrum $T(E)$.  The transmission spectra of the molecule with the different metals
are displayed in Fig.~\ref{f9}.
One of the most visible features of the spectra are the large gap starting 
near the Fermi energy and continuing up to 4-5 eV.  This corresponds to the
HOMO-LUMO gap.  In addition, the spectra are composed of a series of peaks
whose centers correspond to a conducting state of the junction and whose width and height
reflects how strongly that state is coupled to the contacts.  For low voltages,
it is the peaks closest to the Fermi energy that will dominate the transport.
We see in the four cases that we considered that the Fermi energy lies closest
to the HOMO.  Therefore, it is the characteristics of the HOMO that will 
determine many of the features of conduction in these junctions.  Furthermore,
we can read off the zero bias conductance which is given by $G=T(E_f)$ in units
of $2e/h$.  For our junctions, Pt has the highest conductance with $G=0.99$.
After that, we obtain $G=0.29$ for Pd, $G=0.29$ for Au and $G=0.11$ for Ag.

The spatial character of the channels appearing in the transmission spectrum can be
examined with the local density of states (LDOS).  The LDOS can be
extracted directly from the Green's function,
\bg
\rho(\vec{r},E)=-Im Tr(G(\vec{r},E)).
\ed
where the trace is taken over the orbital indices.
Understanding the spatial profile of conduction channels is important for 
engineering molecular devices.  
If we want to affect a particular channel, the LDOS will tells where to focus
our efforts.  For example, if a channel is localized on a sulfur, then we may
want substitute a different atom at that site to get a desired behavior.

In general, we find conducting states of two basic types.  The first are states
based on the molecular bridge.  Typically, these states are extended, conjugated,
$\pi$-bonding states, that span the molecule and are expected to make good
channels to transport electrons.  The other type of state results from the
strong hybridization of the molecule with metallic states of the contacts.

\begin{figure}
\begin{center}
\resizebox{150mm}{!}{\includegraphics{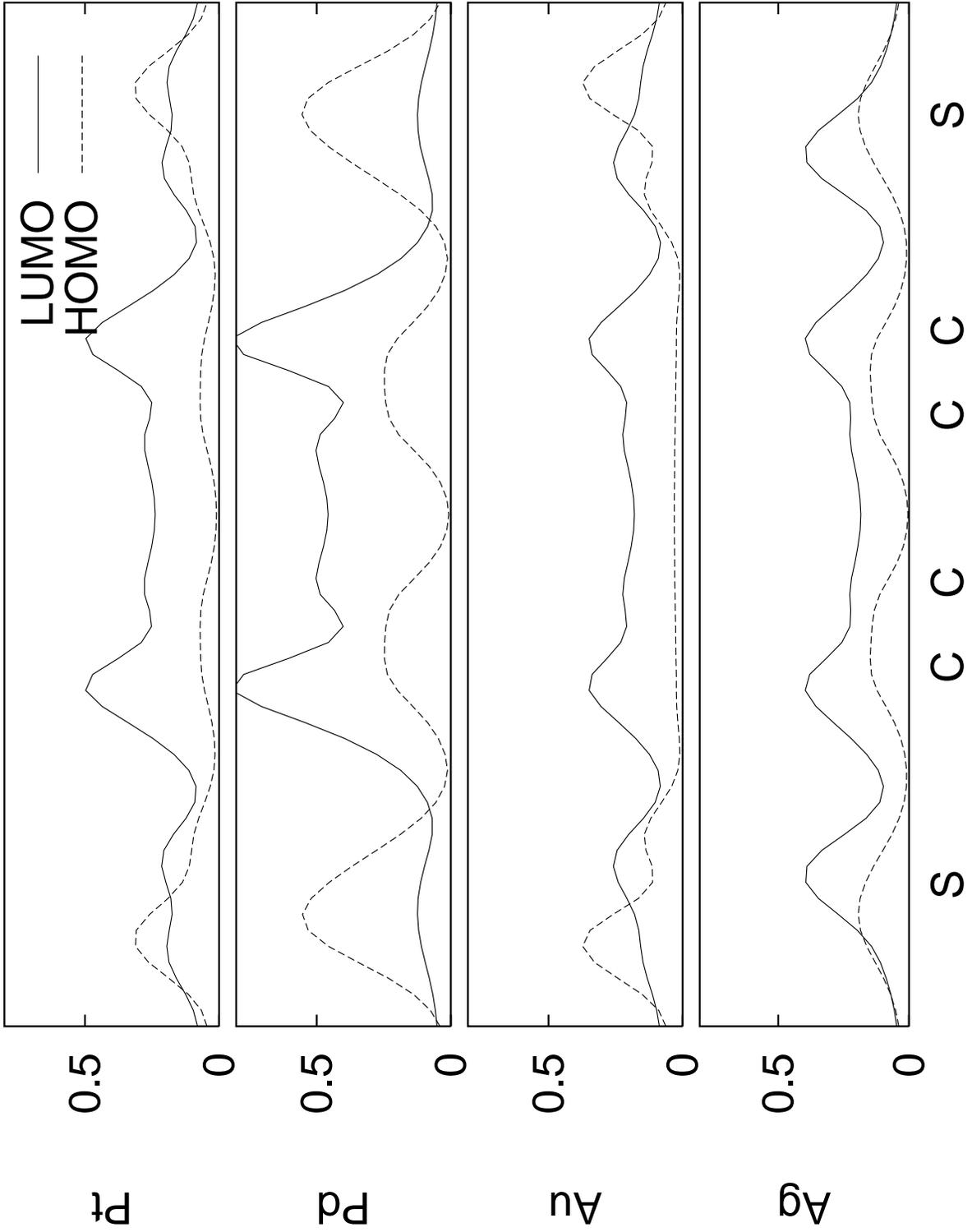}}
\caption{
\label{f10}
Local DOS profile for the HOMO and LUMO for molecular junctions along
the y-axis.
}
\end{center}
\end{figure}

In Fig.~\ref{f10}, we compare the LDOS for the HOMO and LUMO transmission channels
for the four junctions.  As in previous plots, we have averaged over the x and
z directions and show the spatial profile along the axis of the molecule.  In all
cases, the HOMOs appears to have strong weight on the sulfur while the LUMOs are
more distributed across the molecule.  This is important since it is the HOMO which
controls the low bias transport.

\begin{figure}
\begin{center}
\resizebox{150mm}{!}{\includegraphics{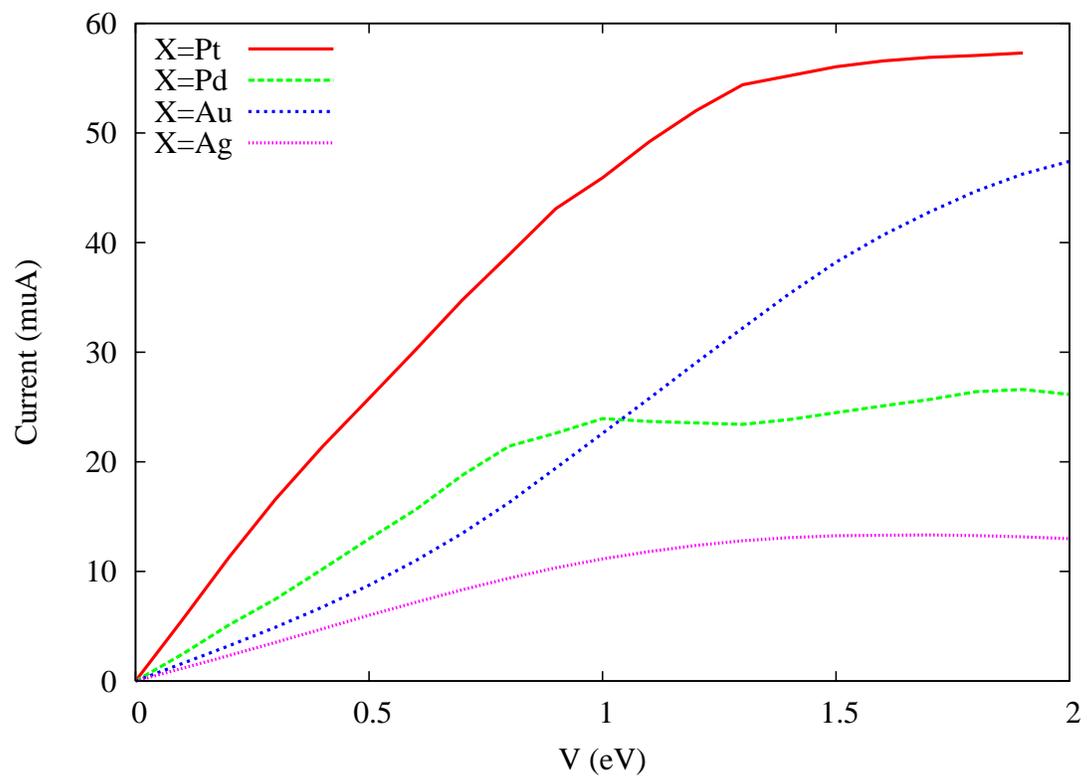}}
\caption{
\label{f11}
(Color online) Comparison of IV characteristics as a function of metal.
}
\end{center}
\end{figure}

The low bias current can be inferred from the equilibrium transmission
spectrum or calculated directly.  In Fig.~\ref{f11}, we show the full IV
characteristics for the different junctions.  Self-consistent calculations were
performed for each bias value.  Notice that flat sections of the IV curve correspond
to gaps in the transmission whereas steep areas signal a new peak entering the
integration window.  For the Au IV curve, there is a small dip in the current for
bias of $1-1.5 V$.  We do not believe this is non-differential resistance
(NDR).  In previous work \cite{cwbjl}, we have seen such bumps, but found improving
the basis set made them disappear.

A comparison of our IV curves with those of Seminario et al. \cite{sem} shows some
significant differences.  First, they have a much larger variation in current flow
with metal. For example at a bias of $1 V$, Pd appears to carry at least two orders of
magnitude more current than Au, while we find very similar currents for Pd and Au.
They also find that Pd has the largest current flow up to biases of $5 V$, while
we find Pt has the largest current flow.

The formalism used by Seminario et al. \cite{sem2}, while in the spirit of many
early calculations, is significantly different from ours especially with respect to
the calculation of the self-energy $\Sigma(E)$.  We believe this explains the difference
between our results.  In their formulation, the surface Green's function $g_s$ that
appears in $\Sigma$ is a constant, diagonal matrix whose elements are the values of
the partial densities of states at $E_f$.  No other electronic structure information
is included in $g_s$.  $\Sigma$, therefore, does not dependent on energy, and there is
no contour integration in their calculations.  There is no recursive method to model
the structure of the semi-infinite bulk contacts.  They use a ``fitting" parameter to
fix the coupling.  Furthermore, the metal clusters are much smaller than what
we use, typically 1-5 Au atoms. 

\begin{figure}
\begin{center}
\resizebox{150mm}{!}{\includegraphics{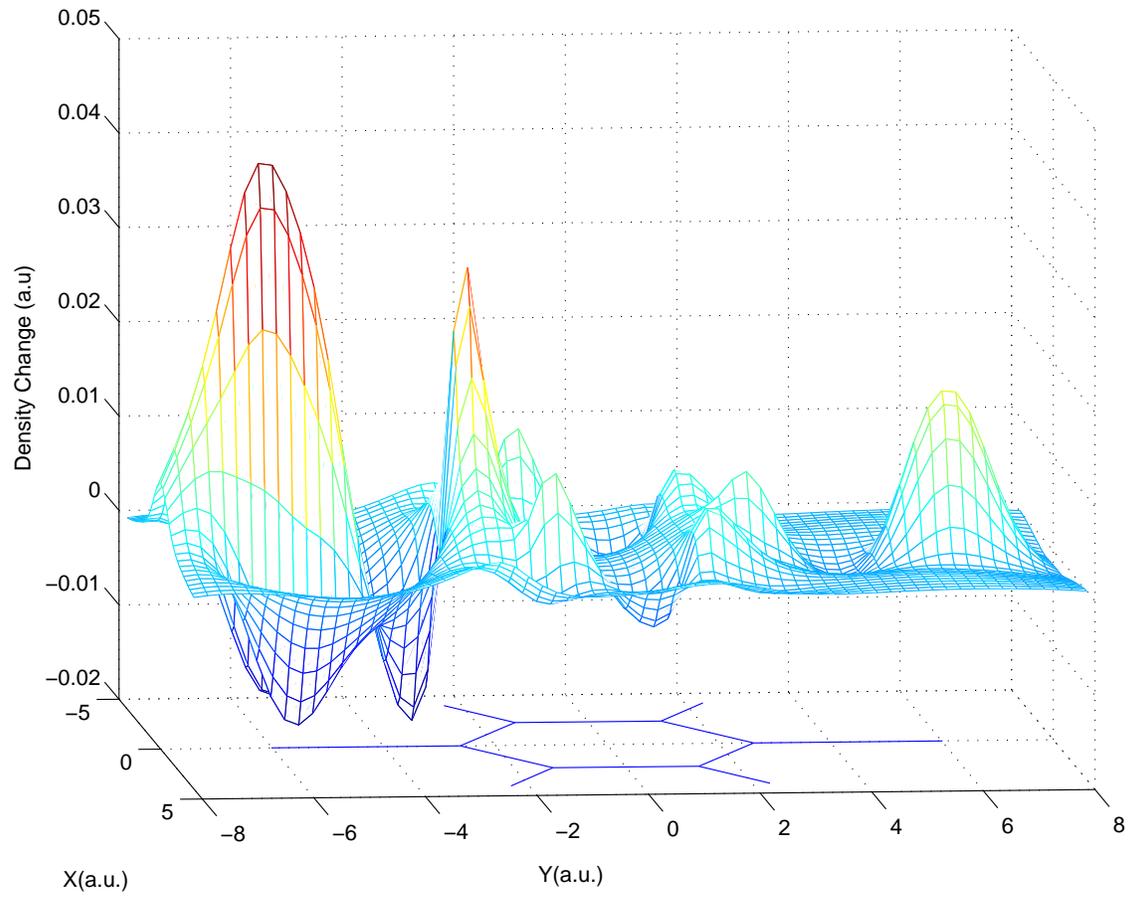}}
\caption{
\label{f12}
(Color online) Spatial profile of charge density redistribution
under bias of 1.8 V for Pd junction.  
}
\end{center}
\end{figure}

\begin{figure}
\begin{center}
\resizebox{150mm}{!}{\includegraphics{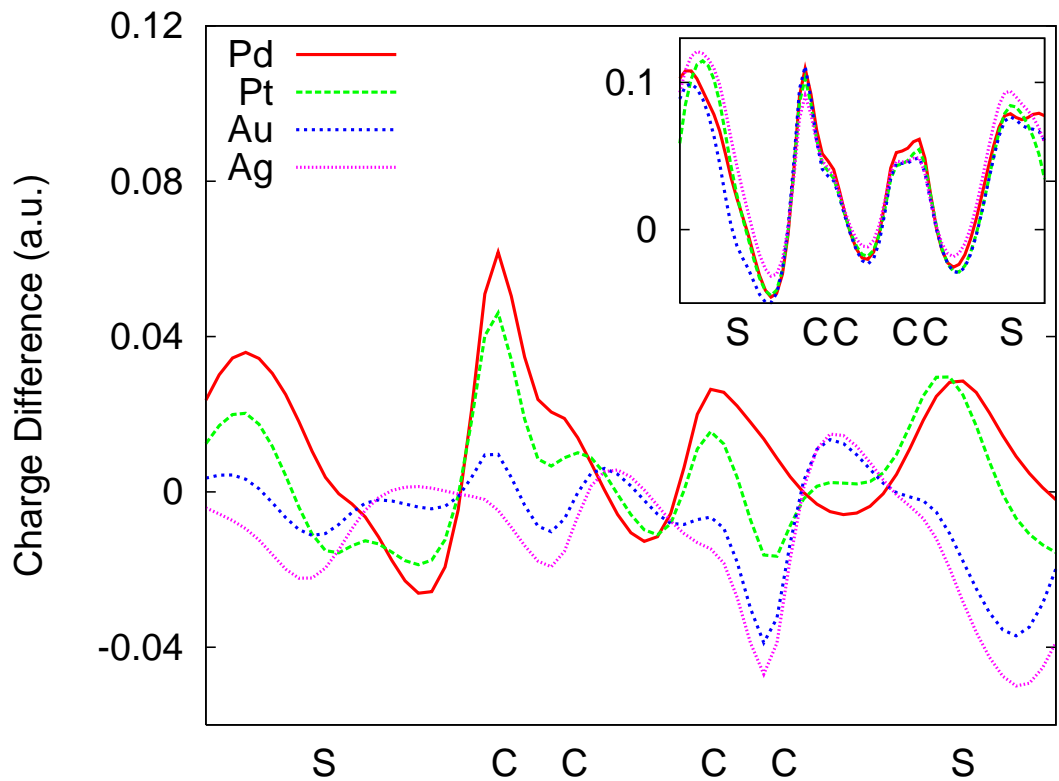}}
\caption{
\label{f13}
(Color online) Comparison of charge density change under bias 1.8 V along the
molecular axis for different junctions.
Main plot is relative to equilibrium junction
while inset is relative to isolated molecule and contacts. 
}
\end{center}
\end{figure}

We can also consider charge density redistribution inside the junction under bias.
It has been pointed out that resistivity dipoles can form due to charge buildup in the
junction \cite{xue031,landibm}.  We find similar effects in the four junctions
we considered.  As an example in Fig.~\ref{f12}, we show the spatial profile
of the charge density redistribution for Pd at a bias of $1.8 V$
The difference relative to the equilibrium density is shown.  We see a large
spike on the left-most carbon and sulfur.  For comparison, we plot in
Fig.~\ref{f13} the density profiles along the y-axis for all the junctions.
Interesting, we see that the largest charge buildup within a junction does
not occur at the same place for the different metals.  For Pd and Pt, the
charge buildup occurs on both sides of C and S, while for Au and Ag, it occurs
on the right.  The inset gives the density redistribution relative to the
isolated molecule and contacts.  This figure should be compared with
Fig.~\ref{f5} to show how the charge buildup affects the original equilibrium charge
transfer.  In particular, we see a reduction on the right of charge accumulation
and an enhancement on the left.  This will affect the respective barrier heights.

\begin{figure}
\begin{center}
\resizebox{150mm}{!}{\includegraphics{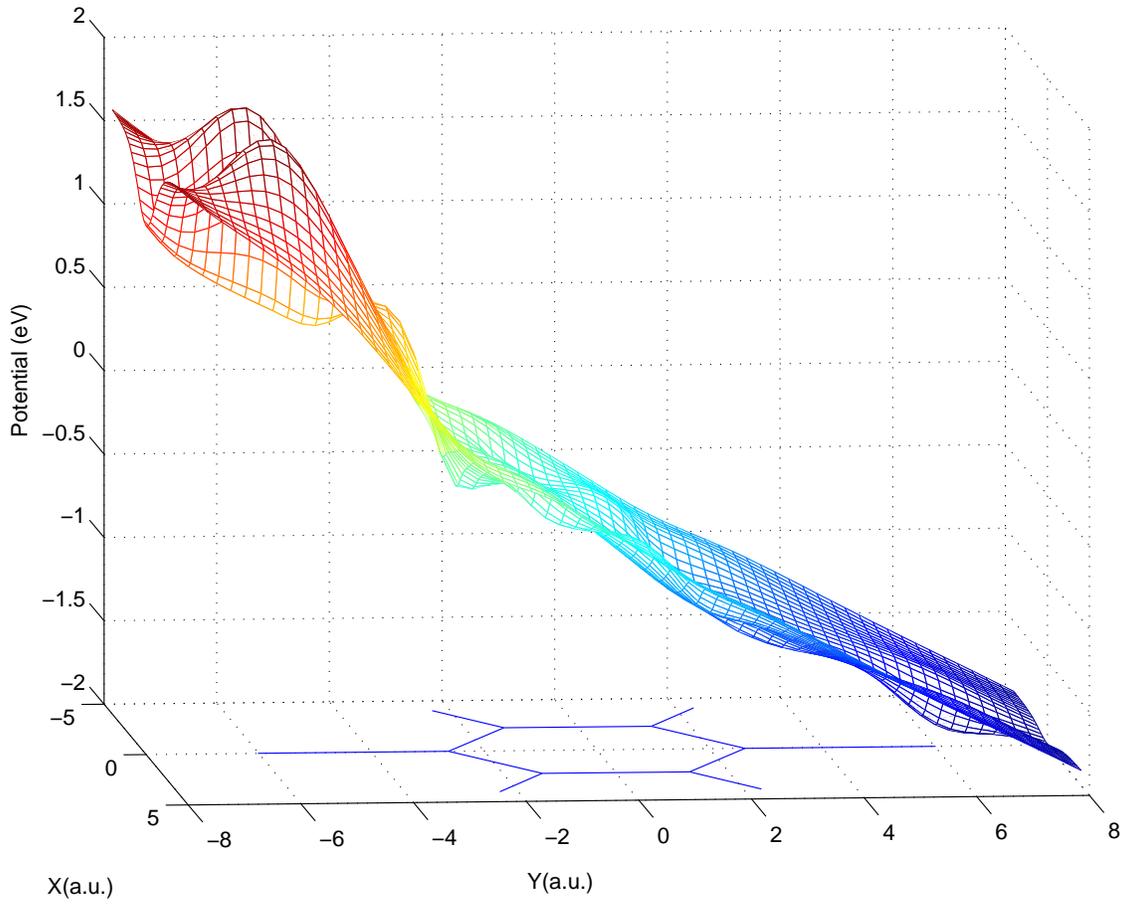}}
\caption{
\label{f14}
(Color online) Spatial profile of electrostatic potential difference under bias
of 1.8 V compared to equilibrium.
}
\end{center}
\end{figure}

\begin{figure}
\begin{center}
\resizebox{150mm}{!}{\includegraphics{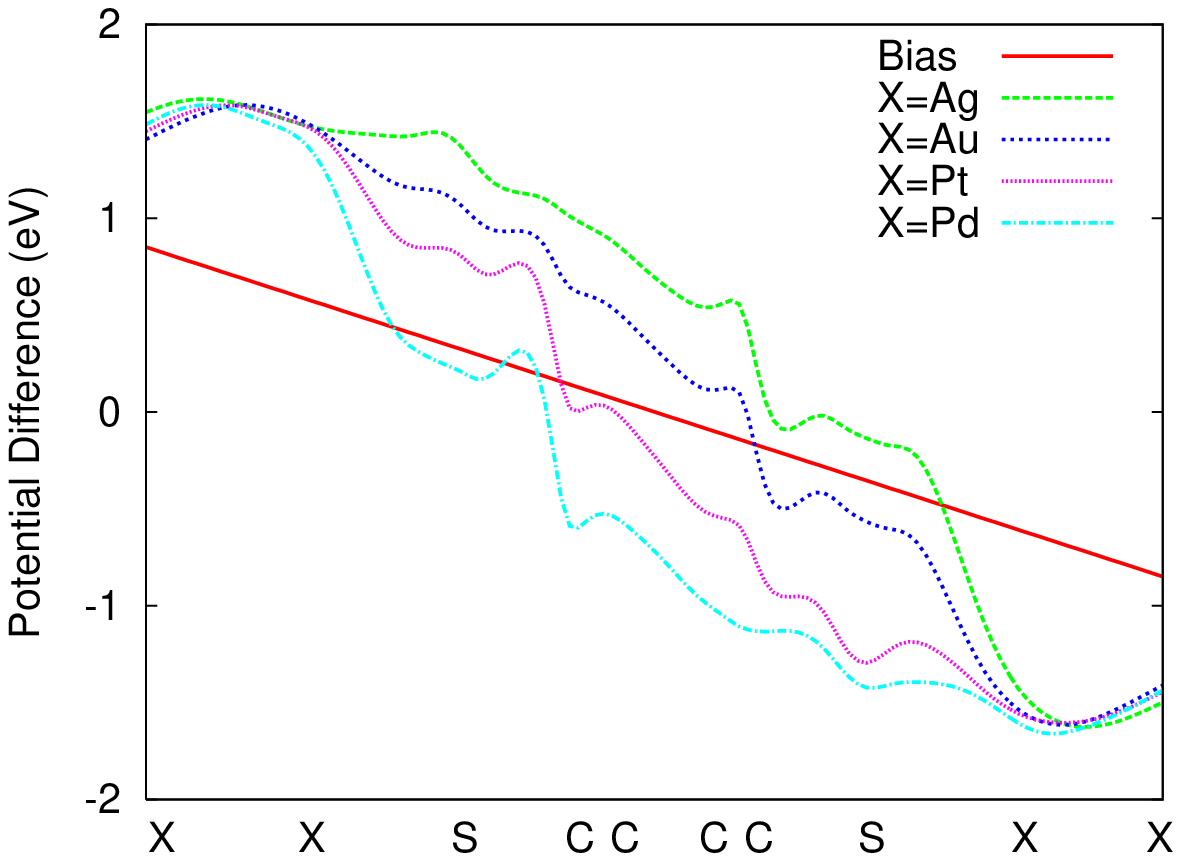}}
\caption{
\label{f15}
(Color online) Comparison of potential differences under bias of 1.8 V
along the molecular axis relative to equilibrium.
}
\end{center}
\end{figure}

Furthermore, we can examine the corresponding change in electrostatic
potential.  Again, for Pd, we show in Fig.~\ref{f14}, the spatial cross
section in the $XY$ plane of the difference between the bias induced potential
and equilibrium.  In Fig.~\ref{f15}, we compare the potential profiles for the
different junctions.  Interesting, there is a spread in the curves for the
different metals inside the junction despite the fact that all the junctions
have the same molecule.  These are effects imposed by the contacts and reflect
the differing polarization inside the molecule as shown in Fig.~\ref{f13}.

\par
\section{Conclusion}
We have performed detailed ab initio calculations of the conduction properties
of PDT connected to Au, Pd, Pt, and Ag electrodes.  We were able to consider the
interplay between equilibrium effects like charge transfer, electrostatics, and
band lineup in the formation of the junction.  The transmission spectra
and the LDOS allowed us to identify the dominant channels for conduction.
In particular, we could consider the spatial distribution of the HOMO and LUMO
and identify where in the molecule these states had greatest weight.

Furthermore, we found that charge transfer in the case of Au and Ag was larger
than with Pd and Pt, resulting in correspondingly larger barriers.  This is
directly reflected in the transport properties where Pt followed by Pd had
the greatest conductance.  Au and Ag on the other hand had the worst.
We were also able to consider in detail the effect of the external bias on 
the redistribution of the charge density and the electrostatic potential.
In particular, we could see a reduction in the magnitude of the charge buildup
resulting from the formation of resistivity dipoles inside the junctions.
Interesting, the location and magnitude of the dipoles varied by junction.

\par
\section{Acknowledgments}
C.W.B. is a civil servant in the Space Technology Division (Mail Stop 230-3), while
J.W.L. is a civil servant in the TI Division (Mail Stop 269-2).\par

\clearpage

\begin{thebibliography}{120}
\bibitem{exp1}{J. Chen, M.A. Reed, A.M. Rawlett and J.M. Tour, Science, 278 (1997) 252}
\bibitem{exp2}{N.P. Guisinger, M.E. Greene. R. Basu, A.S. Baluch and M.C. Hersam Nano Lett 4 (2004) 55}
\bibitem{exp3}{P. Avouris, P.G. Collins and M.S. Arnold, Science 292 (2001) 706}
\bibitem{exp4}{J. Park, A.N. Pasupathy, J.I. Coldsmith, C. Chung, Y. Yaish,J.R. Petta, M. Rinkoski, J.P. Sethna, H.D. Abruna, P.L. McEuen, and D.C. Ralph, Nature, 417 (2002)}
\bibitem{exp5}{W. Ho, X.H. Qiu, and G.V. Nazin, Phys. Rev. Lett. 92 (2004) 206102}
\bibitem{dattaeht}{F. Zahid, M. Paulsson, E. Polizzi, A. Ghosh, L. Siddiqui and S. Datta, J. Chem. Phys. 123 (2005)
064707}
\bibitem{xr1}{Y. Xue, S. Datta, M.A. Ratner, J. Chem. Phys. 115 (2001) 4292}
\bibitem{t2}{P.A. Derosa, J.M. Seminario, J. Phys. Chem. B 105 (2001) 471}
\bibitem{t3}{E.G. Emberly, G. Kirczenow, Phys. Rev. B 64 (2001) 235412}
\bibitem{t4}{P. Damle, A.W. Ghosh, S. Datta, Chem. Phys. 281 (2002) 171}
\bibitem{t5}{M. Di Ventra, S.T. Pantelides, N.D. Lang, Phys. Rev. Lett. 84 (2000) 979}
\bibitem{t6}{L.E. Hall, J.R. Reimers, N.S. Hush, and K. Silverbrook, J. Chem.  Phys. 112 (2000) 1510}
\bibitem{t7}{J. Taylor, H. Guo, J. Wang, Phys. Rev. B 63 (2001) 245407.}
\bibitem{t8}{J. Taylor, M. Brandbyge, K. Stokbro, Phys. Rev. B 68 (2003) 121101}
\bibitem{t9}{H. Ness, S. A. Shevlin, A. J. Fisher, Phys. Rev. B 63 (2001) 125422}
\bibitem{t10}{M. Magoga, C. Joachim, Phys. Rev. B 56 (1997) 4722}
\bibitem{reed}{M.A Reed, C. Zhou, C.J. Miller, T.P. Burgin and J.M. Tour, Science 278 (1997) 252}
\bibitem{h2} {R.H.M Smit, Y. Noat, C. Untiedt, N.D. Lang, M.C. van Hemert and J.M. van Ruitenbeek,
Nature 419 (2002) 906; S. Csonka, A. Halbritter, G. Mihaly, O.I. Shklyarevshii, S. Speller
and H. van Kempen, Phys. Rev. Lett. 93 (2004) 016802}
\bibitem{org}{D. Bong, I. Tam and R. Breslow, JACS 126 (2004) 11796; L.T. Cai, H. Shulason,
J.G. Kushmerick, S.K. Pollack, J. Naciri, R. Shashidhar, D.L. Allara, T.E. Mallouk, and
T.S. Mayer, J. Phys Chem. B 108 (2004) 2827}
\bibitem{sem}{J.M. Seminario, C.E. De La Cruz, and P.A. Derosa, JACS 123 (2001) 5616}
\bibitem {yakrat}{S.N. Yaliraki, M. Kemp, and M. Ratner, JACS, 121, (1999), 3428}
\bibitem {divent}{M. Di Ventra, N.D. Lang, S.T. Pantelides, Chem. Phys, 281, (2002), 189}
\bibitem{xr2}{Y. Xue, S. Datta, M.A. Ratner, Chem. Phys. 281 (2002) 151}
\bibitem{xue031}{Y. Xue, M.A. Ratner, Phys. Rev. B 68 (2003) 115406}
\bibitem{xue032}{Y. Xue, M.A. Ratner, Phys. Rev. B 68 (2003) 115407}
\bibitem{xue04}{Y. Xue, M.A. Ratner, Phys. Rev. B 69 (2004) 085403}
\bibitem{xueth} Y. Xue, Ph.D. thesis, School of Electrical and Computer Engineering,
Purdue University (2000).
\bibitem {cwb1}{C. W. Bauschlicher, J. W. Lawson, A. Ricca, Y. Xue and M. A. Ratner, Chem. Phys. Lett. 388 (2004) 427}
\bibitem{datta}{S. Datta, {\em Electron Transport in Mesoscopic Systems}, Cambridge, 1995 
\bibitem{papa}{D.A. Papaconstantopoulos, {\em Handbook of the Band Structure of Elemental Solids} (Plenum Press, New York, 1986)}
\bibitem{flores}{F. Guinea, C. Tejedor, F. Flores, and E. Louis, Phys. Rev. B 28 (1983) 4397}
\bibitem{cwbar}{C.W. Bauschlicher and A. Ricca, Chem. Phys. Lett. 367, (2003) 90}
\bibitem{cwbxue}{C.W. Bauschlicher and Y. Xue, Chem. Phys. 315 (2005) 293}
\bibitem{g03} {Gaussian 03, Revision B.05,
 M. J. Frisch, G. W. Trucks, H. B. Schlegel, G. E. Scuseria,
 M. A. Robb, J. R. Cheeseman, J. A. Montgomery, Jr., T. Vreven,
 K. N. Kudin, J. C. Burant, J. M. Millam, S. S. Iyengar, J. Tomasi,
 V. Barone, B. Mennucci, M. Cossi, G. Scalmani, N. Rega,
 G. A. Petersson, H. Nakatsuji, M. Hada, M. Ehara, K. Toyota,
 R. Fukuda, J. Hasegawa, M. Ishida, T. Nakajima, Y. Honda, O. Kitao,
 H. Nakai, M. Klene, X. Li, J. E. Knox, H. P. Hratchian, J. B. Cross,
 C. Adamo, J. Jaramillo, R. Gomperts, R. E. Stratmann, O. Yazyev,
 A. J. Austin, R. Cammi, C. Pomelli, J. W. Ochterski, P. Y. Ayala,
 K. Morokuma, G. A. Voth, P. Salvador, J. J. Dannenberg,
 V. G. Zakrzewski, S. Dapprich, A. D. Daniels, M. C. Strain,
 O. Farkas, D. K. Malick, A. D. Rabuck, K. Raghavachari,
 J. B. Foresman, J. V. Ortiz, Q. Cui, A. G. Baboul, S. Clifford,
 J. Cioslowski, B. B. Stefanov, G. Liu, A. Liashenko, P. Piskorz,
 I. Komaromi, R. L. Martin, D. J. Fox, T. Keith, M. A. Al-Laham,
 C. Y. Peng, A. Nanayakkara, M. Challacombe, P. M. W. Gill,
 B. Johnson, W. Chen, M. W. Wong, C. Gonzalez, and J. A. Pople,
 Gaussian, Inc., Pittsburgh PA, 2003.}
\bibitem {becke}{A.D. Becke  Phys Rev A 38 (1988) 3098}
\bibitem {pw91}{J.P. Perdew, Y. Wang, Phys Rev B  45 (1991) 13244 }
Cambridge, 1995)}
\bibitem {cwbjl}{C. W. Bauschlicher, J. W. Lawson, Chem. Phys., 324, (2006), 647}
\bibitem{sem2}{P. Derosa and J. Seminario, J.Phys.Chem. B, 105, (2001) p471}
\bibitem{landibm}{R. Landauer, IBM J. Res. Dev 1 (1957) 223, 32 (1988) 306}
\end{thebibliography}
\end {document}